\documentclass[runningheads]{llncs}

\usepackage{amsmath,amsfonts,amssymb}
\usepackage{graphicx}
\usepackage{subcaption}
\usepackage[dvipsnames]{xcolor}
\usepackage[symbol]{footmisc}
\usepackage{makecell}
\usepackage{subfiles}

\renewcommand{\thefootnote}{\fnsymbol{footnote}}

\graphicspath{ {images/} }

\newif\ifdraft
\draftfalse 

\ifdraft
\def\H#1{\textcolor{Aquamarine}{#1}}
\def\Hc#1{\textcolor{Aquamarine}{\textit{\textsf{ \small [HB: #1]}}}}  
\def\Hd#1{\textcolor{purple}{\textit{[deleted: #1]}}}  

\def\D#1{\textcolor{Orange}{#1}}
\def\Df#1{\textcolor{red}{#1}}
\def\Dc#1{\textcolor{Orange}{\textit{\textsf{ \small [DL: #1]}}}}  
\def\Dd#1{\textcolor{purple}{\textit{[deleted: #1]}}}  

\def\P#1{\textcolor{blue}{#1}}
\def\Pc#1{\textcolor{blue}{\textit{\textsf{ \small [PS: #1]}}}}  
\def\Pd#1{\textcolor{purple}{\textit{[deleted: #1]}}}  

\else
\def\H#1{#1}
\def\Hc#1{}  
\def\Hd#1{}  

\def\D#1{#1}
\def\Df#1{#1}
\def\Dc#1{}  
\def\Dd#1{}  

\def\P#1{#1}
\def\Pc#1{}  
\def\Pd#1{}  

\fi

\begin{document}
%
\title{Projective Skip-Connections for Segmentation Along a Subset of Dimensions in Retinal OCT}
\titlerunning{Projective Skip-Connections}
\authorrunning{D. Lachinov et al.}
%
\author{Dmitrii Lachinov\inst{1,2} \and 
Philipp Seeböck\inst{1} \and 
Julia Mai\inst{1,2} \and 
Felix Goldbach\inst{1} \and 
Ursula Schmidt-Erfurth\inst{1} \and  
Hrvoje Bogunovic\inst{1,2}  
}
%
%
\institute{Department of Ophthalmology and Optometry,\\ Medical University of Vienna, Austria \and
Christian Doppler Laboratory for Artificial Intelligence in Retina,\\
Department of Ophthalmology and Optometry, Medical University of Vienna, Austria
}
\maketitle              
\begin{abstract}
In medical imaging, there are clinically relevant segmentation tasks where the output mask is a projection to a subset of input image dimensions.
In this work, we propose a novel convolutional neural network architecture that can effectively learn to produce a lower-dimensional segmentation mask than the input image. The network restores encoded representation only in a subset of input spatial dimensions and keeps the representation unchanged in the others. The newly proposed projective skip-connections allow linking the encoder and decoder in a UNet-like structure.
We evaluated the proposed method on two clinically relevant tasks in retinal Optical Coherence Tomography (OCT): geographic atrophy and retinal blood vessel segmentation.
The proposed method outperformed the current state-of-the-art approaches on all the OCT datasets used, consisting of 3D volumes and corresponding 2D en-face masks.
The proposed architecture fills the methodological gap between image classification and $N$D image segmentation.
\end{abstract}
\section{Introduction}

%

The field of medical image segmentation is dominated by neural network based solutions. The convolution neural networks (CNNs), notably U-Net \cite{unet} and its variants, demonstrate state-of-the-art performance on a variety of medical benchmarks like BraTS \cite{brats}, LiTS \cite{lits}, REFUGE \cite{refuge}, CHAOS \cite{chaos} and ISIC \cite{isic}. Most of these benchmarks focus on the problem of segmentation, either 2D or 3D, \H{where the input image and the output segmentation mask are of the same dimension}. However, a few medical protocols \Df{like OCT for retina \cite{IPN,selected_dimensions}, OCT and Ultrasound for skin \cite{skin_ultrasound,skin_oct}, Intravascular ultrasound (IVUS) pullback images for vasculature \cite{vasc}, CT for diaphragm analysis \cite{diaphragm} and online tumor tracking \cite{tracking} require the segmentation to be performed on the image projection,} \H{resulting in the output segmentation of lower spatial dimension than the input image}. It introduces the problem of dimensionality reduction into the segmentation, for example, segmenting a flat 2D en-face structures on the projection of a 3D volumetric image.

This scenario has so far not been sufficiently explored in the literature and it is currently not clear what CNN architectures are the most suitable for this task. Here, we propose a new \P{approach} for $N\text{D}\rightarrow M\text{D}$ segmentation\D{, where $M < N$ }, \D{and evaluate} it on two clinically relevant tasks of 2D Geographic Atrophy (GA) and Blood Vessels segmentation in 3D optical coherence tomography (OCT) retinal images. 

\subsection{Clinical background}
\begin{figure}[t]
\centering
\begin{subfigure}{.2\textwidth}
  \centering
  \includegraphics[width=\linewidth]{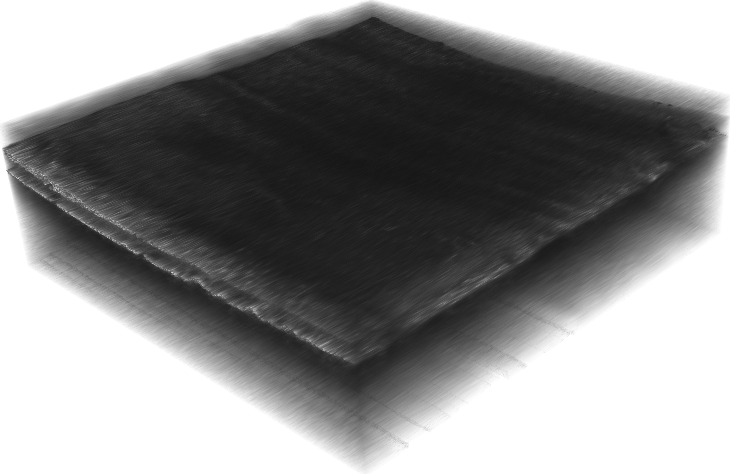}
  \caption{}
\end{subfigure}%
\begin{subfigure}{.2\textwidth}
  \centering
  \includegraphics[width=\linewidth,height=0.7\linewidth]{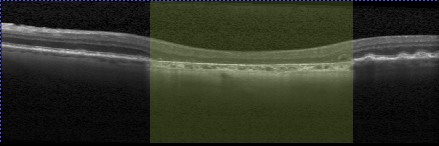}
  \caption{}
\end{subfigure}%
\begin{subfigure}{.2\textwidth}
  \centering
  \includegraphics[width=\linewidth,height=0.7\linewidth]{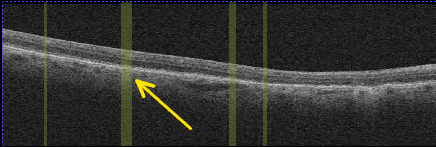}
  \caption{}
\end{subfigure}%
\begin{subfigure}{.2\textwidth}
  \centering
  \includegraphics[width=\linewidth]{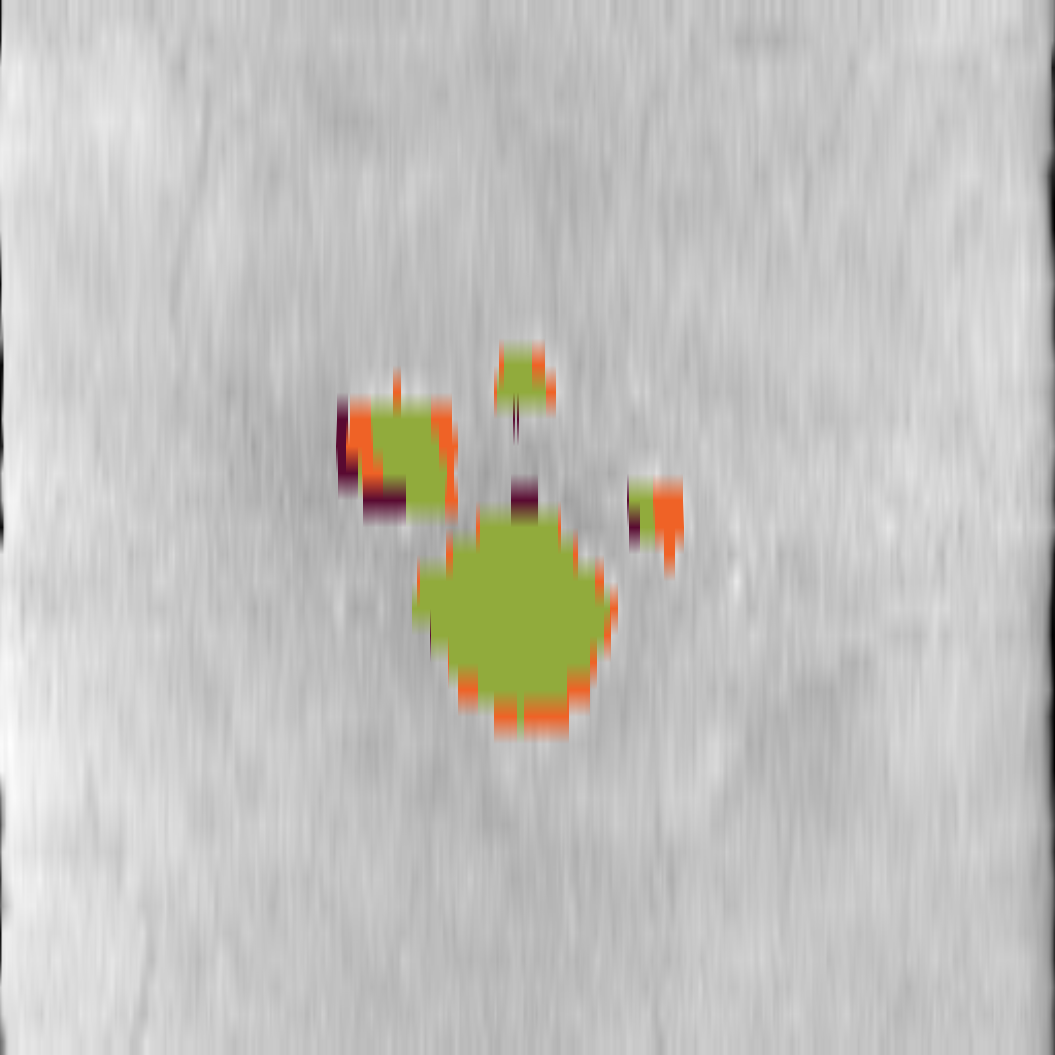}
  \caption{}
  \label{fig:ga_example:d}
\end{subfigure}%
\begin{subfigure}{.2\textwidth}
  \centering
  \includegraphics[width=\linewidth]{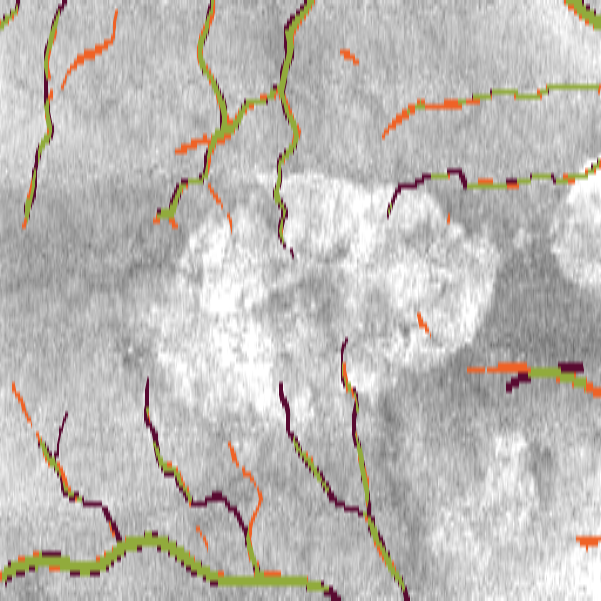}
  \caption{}
  \label{fig:vess_example:e}
\end{subfigure}
\caption{(a) 3D retinal OCT volume. \P{Ground truth annotations are shown in 2D cross-sectional slices for (b) GA and (c) blood vessels in yellow. Predictions of the proposed method are visualized in en-face projection images for (d) GA and (e) blood vessels.} Green region - true positives, orange - false positives, dark red - false negatives.}
\label{fig:ga_example}
\end{figure}

\paragraph{Geographic atrophy (GA)} is an advanced form of age-related macular degeneration (AMD). It corresponds to localized irreversible and progressive loss of retinal photoreceptors and leads to a devastating visual impairment. 3D optical coherence tomography (OCT) is a gold standard modality for retinal examination in ophthalmology. In OCT, GA is characterized by hypertransmission of OCT signal below the \P{retina}. \H{As GA essentially denotes a loss of tissue, it does not have ``thickness''}, and it is hence delineated as a 2D en-face area (Fig. \ref{fig:ga_example}).

\paragraph{Retinal blood vessels} provide oxygen and nutrition to the retinal tissues. Retinal vessels are typically examined using 2D fundus photographs where they can be seen as dark lines. In OCT volumes, retinal blood vessels can be detected in individual slices (B-scans) as interconnected morphological regions, dropping shadows on the underlying retinal layer structures (Fig. \ref{fig:ga_example}).

\subsection{Related work}
Recently, several methods dealing with dimensionality reduction in segmentation tasks have been proposed~\cite{selected_dimensions,IPN,deep_voting}. 
Liefers et al. \cite{selected_dimensions} introduced a fully convolutional neural network to address the problem of $N\text{D}\rightarrow M\text{D}$ segmentation.
The authors evaluated their method on \P{$2D\rightarrow 1D$} tasks of Geographic Atrophy (GA) and Retinal Layer segmentation. \P{In contrast to our method, t}heir approach is limited by the fixed size of the input image and shortcut networks that are prone to overfitting. \P{This is caused by the full compression of the image in 2nd dimension, leading to a large receptive field but completely removing the local context.}

Recently, Li et al. \cite{IPN} proposed an image projection network IPN designed for 3D-to-2D image segmentation, without any explicit shortcuts or skip-connections. Pooling is only performed for a subset of dimensions, resulting in a highly anisotropic receptive field. This limits the amount of context the network can use for segmentation.

Ji et al. \cite{deep_voting} proposed a deep voting model for GA segmentation. First, multiple fully-connected classifiers are trained \H{on axial depth scans (A-scans)}, producing a single probability value for each A-scan (1D-to-0D reduction). During the inference phase, the predictions of the trained classifiers are concatenated to form the final output. This approach doesn't account for neighboring AScans, thus lacks spatial context, and requires additional postprocessing.

\D{Existing} approaches suffer from two main limitations: On the one hand, they are designed to handle images of a fixed size, making patch based training difficult. This is of particular relevance as it has been shown that patch based training improves overall generalization as it serves as additional data augmentation~\cite{alexnet}. On the other hand, existing $N\text{D}\rightarrow M\text{D}$ methods have a large receptive field due to a high number of pooling operations or large pooling kernel sizes, that fail to capture local context. To illustrate this issue, we conducted a compared receptive fields of state-of-the-art segmentation~\cite{Jiang2019,Zhao2019,McKinley2019,Han2017AutomaticLL,Vorontsov2018LiverLS,Bi2017AutomaticLL} and classification methods~\cite{Matsunaga2017ImageCO,Daz2017IncorporatingTK,Menegola2017RECODTA} for medical imaging across different benchmarks~\cite{brats,lits,isic} with popular architectures designed for natural images classification \cite{resnet,densenet}\H{(Fig 1-2 in the supplement)}. We observed that current state-of-the art methods for segmentation in a subset of input dimensions \cite{selected_dimensions,IPN} \textit{fall out of the cluster with methods designed for medical tasks}, and have a receptive field comparable to the networks designed for natural images classification. However, the amount of training data available for medical tasks differs by the orders of magnitude from the natural images data. \P{Alternatively, the $N\text{D}\rightarrow M\text{D}$ segmentation task can be solved using $N\text{D}\rightarrow N\text{D}$ methods. It can be attempted by either first projecting the input image to the output $M\text{D}$ space, which looses context, or by running $N\text{D}$ segmentation first, which is memory demanding, requires additional postprocessing and is not effective.}

Our approach is explicitly designed to overcome these limitations. We propose a segmentation network that can handle arbitrary sized input and is on par with state-of-the-art $N\text{D}\rightarrow N\text{D}$ medical segmentation methods in terms of receptive field size.

\subsection{Contribution}
In this paper we introduce a novel CNN architecture for $N\text{D}\rightarrow M\text{D}$ segmentation. The proposed approach has a decoder that operates in the same dimensionality space as the encoder, and restores the compressed representation \P{of the bottleneck layer} only in a subset of the input dimensions. \P{The contribution of this work is threefold: (1) We propose projective skip-connections addressing the general problem of segmentation in a subset of dimensions; (2) We provide clear definitions and instructions on how to reuse the method for arbitrary input and output dimensions; (3) We perform an extensive evaluation with three different datasets on the task of Geographic Atrophy and Retinal Blood Vessel $3D\rightarrow 2D$ segmentation in retinal OCTs.}
\P{The results demonstrate that the proposed method clearly outperforms the state-of-the-art.}

\section{Method} 


Let the image $I \in \mathbb{R}^{\prod_{d=1}^{N}{n_d}}$, where $N$ is the number of dimensions, and $n_d$ is the size of the image in the corresponding dimension $d$. We want to find such function $f$ that $f:I \longrightarrow  O$, where $O \in \mathbb{R}^{\prod_{d =1}^{M}{n_d}}$ is output segmentation and $M$ is its dimensionality. We focus on the case where $M\leq N$ with dimensions $d\le M$ being target dimensions, where we perform segmentation; and $M<d\le N$ being reducible dimensions.

We parameterize $f$ by a convolutional neural network. The CNN architecture we propose to use for the segmentation along dimensions $M$ follows a U-Net \cite{unet,3dunet} architecture. The encoder consists of multiple blocks that sequentially process and downsample the input volume. Unlike the other methods, we don't perform the global pooling \D{in the network bottleneck} in the dimensions outside of $M$. \H{Instead,} we freeze the size of these dimensions and propagate the features through the decoder into the final classification layer. \P{The} decoder restores the input resolution only across those \P{target} dimensions $M$ where we perform the segmentation. The decoder keeps feature maps in the remaining \P{reducible} dimensions $M < d \le N $ compressed. Since the sizes of encoder and decoder feature maps do not match, \P{we propose \textit{projective skip-connections} to link them.}

\Df{At the last stage of the network}, each location of the output $M$D image has a corresponding feature tensor of compressed representation in $(N-M)$D \Df{(refer to Fig.~\ref{fig:arch})}. We \Dd{first }perform Global Average Pooling (GAP) \Dd{in these compressed dimensions }\Dd{to reduce the feature tensor size to 1. Then, we use}and a regular convolution to obtain $M$D logits.

\subsection{\H{Projective} skip-connections}
\begin{figure}[t]
\centering
\includegraphics[width=\textwidth]{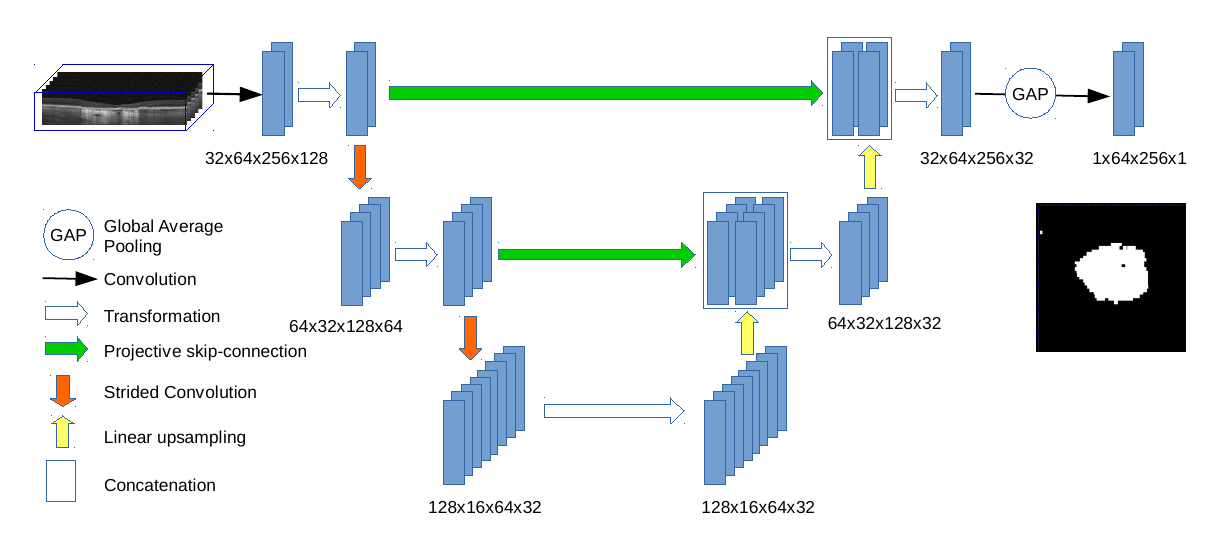}
\caption{The proposed CNN architecture solving $ND\rightarrow MD$ segmentation problem, where $N=3$ and $M=2$. \emph{Transformation} can be any sort of operation, residual blocks in our case. The input is a 3D crop from the SD-OCT image. The final feature map is averaged (GAP) in dimensions that are not in $M$, only last dimension in our case. Then the feature map is processed with final convolution, that reduces the number of channels from 32 to 1. The output is a 2D en-face segmentation mask. The network in this figure has a configuration of $l=3$, $C = \{2, 4, 8\}$, and $B=\{1,1,1\}.$
}
\label{fig:arch}
\end{figure}

The purpose of projective skip-connections is to compress the reducible dimensions $M < d \le N $ to the size of bottleneck and leave the target dimensions $d \le M$ unchanged. \Df{In contrast to GAP, we don't completely reduce the target dimensions to size of 1 (Section \ref{sec:baselines}).} \Df{Instead,} we use average pooling with varying kernel size \Df{(Fig. \ref{fig:arch})}. Keeping in mind that the feature map size along dimension $d$ generated by \P{our proposed} network $f$ of depth $l$ with input image $I$ at the level $j$ of the encoder will be $\frac{n_d}{2^{j-1}}$, the feature map size at the level $j$ of the \textbf{decoder} will be $n_d^j$ (\ref{1stequation}).\\
\begin{minipage}{\textwidth}
\begin{minipage}{0.5\textwidth}
\begin{equation}
\label{1stequation}
n_d^j = \begin{cases} \frac{n_d}{2^{j-1}} , & \mbox{if } d \le M \\ \frac{n_d}{2^{l-1}}, & \mbox{otherwise.} \end{cases} 
\end{equation}
\end{minipage}
\begin{minipage}{0.5\textwidth}
\begin{equation}
\label{2ndequation}
k_d^j = \begin{cases} 1 , & \mbox{if } d \le M \\ 2^{l-j}, & \mbox{otherwise,} \end{cases}
\end{equation}
\end{minipage}
\end{minipage}
We propose to perform the average pooling of the encoder feature maps with kernel size and stride $k_d^j$ (\ref{2ndequation}), where $d$ is the corresponding dimension, and $j$ is the corresponding layer of the network. After the average pooling is performed, we concatenate the encoder and decoder feature maps.

The proposed CNN can be explicitly described with the depth or the number of total network levels $l$, the number of channels per convolution for each level $C = \{{c_0}\times 2^{i-1}: 1\leq i \leq l\}$, and the number of residual blocks at each level $B = \{b_i: i \leq l\}$.
The schematic representation of the proposed approach is shown in Fig.~\ref{fig:arch}. The network has a configuration of $l=3$, $C = \{2, 4, 8\}$ and $B=\{1,1,1\}$. The input is a 3D volume \P{patch} $I\in \mathbb{R}^{64\times128\times256}$, so $N = 3$. We are interested in the en-face segmentation, so $M = 2$.
In the extreme case where $M = N$, the proposed CNN is equivalent to a U-Net architecture with residual blocks. If $M=0$, the proposed CNN is equivalent to \P{N-D }ResNet with skip-connections in the last residual blocks.

\section{Experiments}
\subsection{Data sets}
We use a rich collection of three medical datasets with annotated retinal OCT volumes for our experiments (Table \ref{tab_datasets_stats}). All our datasets consist of volumetric 3D OCT images and corresponding \H{en-face} 2D annotations \H{(Fig \ref{fig:ga_example})}, meaning that the used datasets represent $3D \rightarrow 2D$ segmentation problems.

\begin{table}[t]
\centering
\caption{Datasets description}
\label{tab_datasets_stats}
\centering
\resizebox{\textwidth}{!}{\begin{tabular}{|c|c|c|c|c||c|c|c|c|c|c|}
\hline
\thead{Dataset} & Device & Volumes & Eyes & Patients & \thead{Depth,\\mm} & \thead{En-face\\area,$\text{mm}^2$} & \thead{Depth\\spacing, $\mu\text{m}$} & \thead{En-face\\Spacing,$\mu\text{m}^2$} & \thead{ Depth\\resolution, px} & \thead{En-face\\resolution, $\text{px}^2$} \\
\hline
GA 1 & Spectralis & 192 & 70 & 37 & 1.92 & $6.68\times 6.86$ & $3.87$ &  $156.46\times7.91$ & 496 & $43\times867$ \\
\hline
GA 2 & Spectralis & 260 & 147 & 147 & 1.92 & $6.02\times6.03$ & $3.87$ & $122.95\times11.16$ & 496 & $49\times540$ \\
\hline
Vessel 1 & TopconDR & 1 & 1 & 1 & 2.3 & $6\times6$ & $2.6$ & $46.87\times11.72$ & 885 & $128\times512$ \\
         & Cirrus & 40 & 40 & 40 & 2 & $6.05\times6.01$ & $1.96$ & $47.24\times11.74$ & 1024 & $128\times512$ \\
         & Spectralis & 43 & 43 & 43 & 1.92 & $5.88\times5.82$ & $3.87$ & $60.57\times5.68$ & 496 & $97\times1024$ \\
         & Topcon2000 & 9 & 9 & 9 & 2.3 & $6\times6$ & $2.6$ & $46.88\times11.72$ & 885 & $128\times512$ \\
         & Topcon-Triton & 32 & 32 & 32 & 2.57 & $7\times7$ & $2.59$ & $27.34\times13.67$ & 992 & $256\times512$ \\
\hline
\end{tabular}}
\end{table}

The following are the three datasets. \textbf{Dataset `GA1':} The scans were taken at multiple time points from an observational longitudinal study of natural GA progression. Annotations of GA were performed by retinal experts on 2D Fundus Auto Fluorescence (FAF) images and transferred by image registration to the OCT, resulting in 2D en-face OCT annotations. \textbf{Dataset `GA2':} With the OCT scans originating from a clinical trial that has en-face annotations of GA that were annotated directly on the OCT BScans by a retinal expert. \textbf{Dataset `Vessel~1':} A \H{diverse} set of OCT scans \H{across device manufacturers} \H{of patients with AMD} where the retinal blood vessels were directly annotated on OCT en-face projections by retinal experts. 

\subsection{Baseline methods}\label{sec:baselines}
We exhaustively compared our approach against multiple $N\text{D}\rightarrow N\text{D}$ baselines: 
\emph{UNet 3D} \cite{3dunet} operating on OCT volume;
\emph{UNet 2D} \cite{unet} operating on OCT volume projections;
\emph{UNet++} \cite{unetpp} operating on OCT volume projections. And $N\text{D}~\rightarrow~M\text{D}$ methods: 
\emph{SD} or Selected Dimensions \cite{selected_dimensions};
\emph{FCN} or Fully Convolutional Network, used as a baseline in~\cite{selected_dimensions};
\emph{IPN} or Image Projection Network \cite{IPN};

\textit{Ablation \P{ experiment}:} \emph{3D2D} is a variation of the proposed method with 3D encoder and 2D decoder networks. The bottleneck and skip-connections perform global pooling in the dimensions $d \ge M$. 
\P{This ablation experiment was included to study} the effect of propagating deep features to the last stages of the network.

\subsection{Training Details}
As a part of preprocessing, we flattened the volume along the Bruch's Membrane, and cropped a 3D retina region with the size of 128 pixels along AScan (vertical direction) and \H{keep the} full size in the rest of dimensions. We resampled the images to have $119.105\times5.671 \mu\text{m}^2$ en-face spacing. Before processing the data, we performed z-score normalization in cross-sectional plane. \Df{For generating OCT projections, we employed the algorithm introduced by Chen et al. \cite{rsvp}}.

\textit{For the task of Geographic Atrophy segmentation}, the \emph{proposed} method has the following configuration: $l=4$, $B=\{1,1,1,1\}$, $C=\{32,64,128,256\}$.
We use residual blocks \cite{resnet} with 3D convolutions with kernel size of 3. Instead of Batch Normalization \cite{batchnorm}, we employ Instance Normalization \cite{instancenorm}, due to small batch size.
The \emph{3D2D} network has the same configuration as the proposed method. \emph{SD} and \emph{FCN} are reproduced as in the paper \cite{selected_dimensions}, but with the AScan size equals to 128 and 32 channels in the first convolution. \emph{IPN} network is implemented as in the paper~\cite{IPN} for $3D\rightarrow 2D$ case with 32 channels in the first convolution. \emph{UNet 2D} is reproduced as in the paper \cite{unet}, but with residual blocks. \emph{UNet 3D}\cite{3dunet} and \emph{UNet++}\cite{unetpp} were reproduced by following the corresponding papers. \emph{UNet 3D} output masks were converted to 2D \Df{by pooling the reducible dimensions }and thresholding. \textit{For Retinal Vessel segmentation}, we used the networks of the same configuration, but we set the number of channels in the first layer equal to 4.

The models were trained with Adam \cite{adam} optimizer for $3\cdot 10^4$ and $10^4$ iterations with weight decay of $10^{-5}$, learning rate of $10^{-3}$ and decaying with a factor of 10 at iteration $2\cdot 10^4$  and $6\cdot 10^3$ for GA and blood vessels segmentation respectively. 
For GA segmentation, the batch size was set to 8 and patch size to $64\times 256\times 64$. For the \emph{SD} and \emph{FCN} models we used patch size of $64\times 512\times 512$ due to architecture features described in the paper \cite{selected_dimensions}.
For blood vessels segmentation, the batch size was set to 8 and patch size was $32\times 128\times 256$. The optimizer was chosen empirically for each model.
We used Dice score as a loss function for both tasks.


\subsection{Evaluation Details}
For all three datasets (\textit{GA1}, \textit{GA2}, \textit{Vessel~1}) we conducted a 5-fold cross-validation for evaluation. The splits were made on patient level with stratification by baseline GA size for \textit{GA} datasets and by device manufacturer for \textit{Vessel~1} dataset. \D{The results marked as validation are average of all samples from all the validation splits.} The \textit{Dice score} and 95th percentile of Hausdorff distance averaged across scans (Dice and HD95) were used. \P{To test for significant differences between the proposed method and baselines, we conducted two-sided Wilcoxon signed-rank tests, using $\alpha=0.05$.}
In addition to cross-validation results, we report performances on hold out (\textit{external}) datasets: models trained on \textit{GA1} were evaluated on \textit{GA2} serving as an external test set and vice versa.

\section{Results and Discussion}

\begin{figure}[th]
\centering
\begin{subfigure}{0.49\textwidth}
  \centering
  \includegraphics[width=\linewidth]{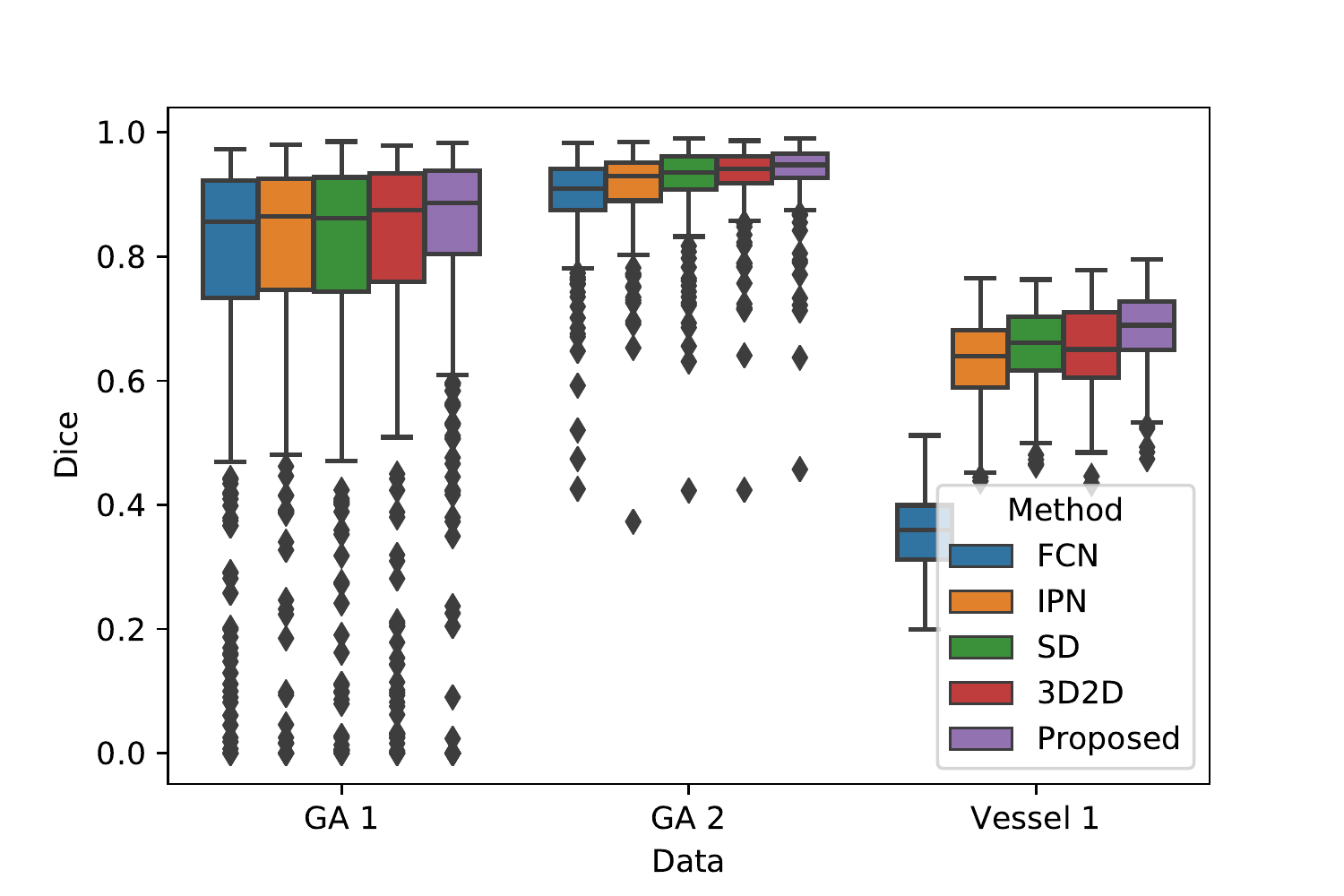}
  \caption{}
  \label{fig:boxplot3}
\end{subfigure}%
\begin{subfigure}{0.49\textwidth}
  \centering
  \includegraphics[width=\linewidth]{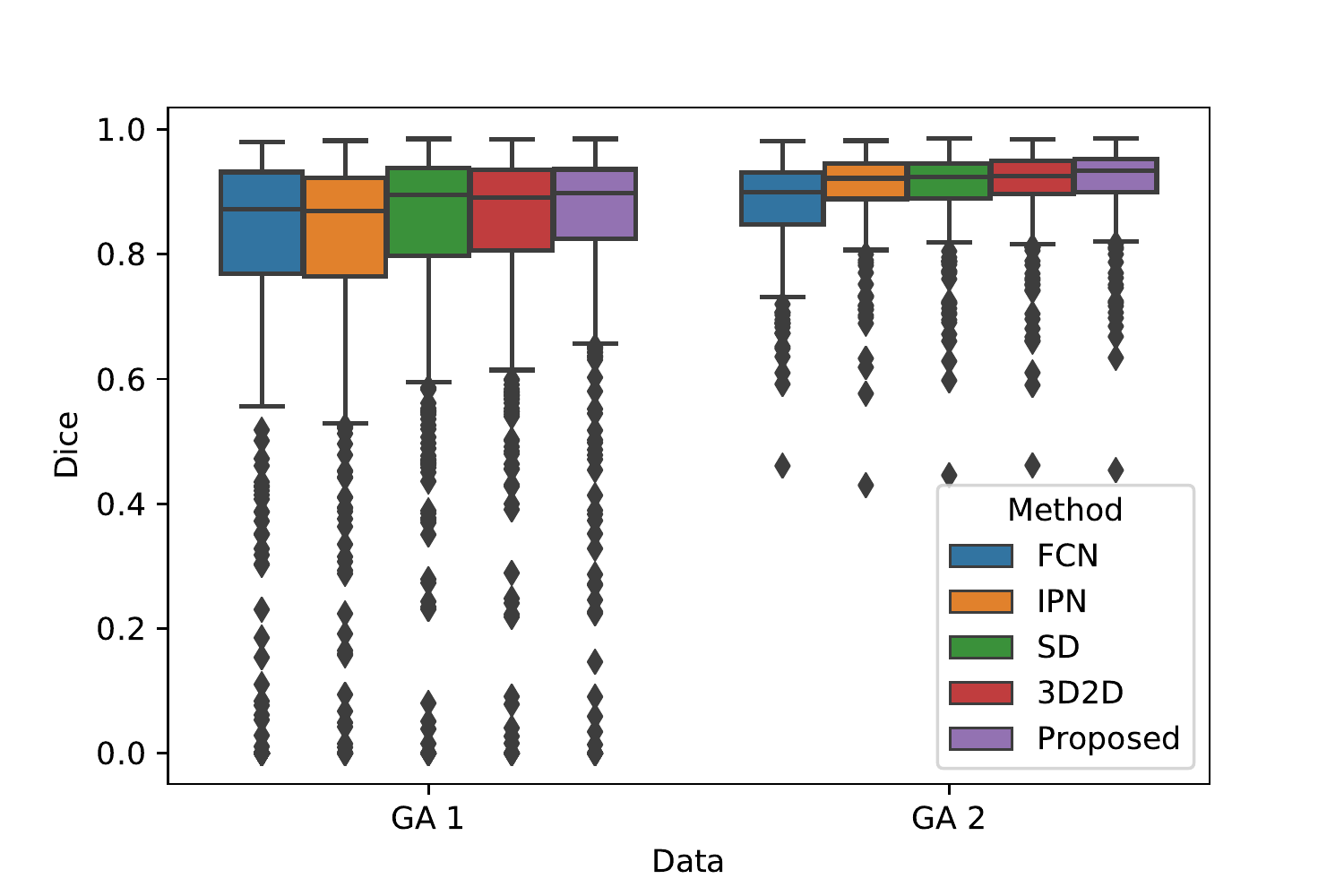}
  \caption{}
  \label{fig:boxplot_cross}
\end{subfigure}
\caption{Segmentation performance evaluated with (a) cross-validation and (b) external test set.}
\end{figure}

\textit{Cross-validation} results are reported in the Table \ref{tab2}. The qualitative examples of segmentations can be found in Supplementary materials. For all the tasks, the proposed methods achieved a significant improvement in both the mean \emph{Dice} and \emph{Hausdorff} distance. \P{This improvement is also reflected in the boxplots, showing higher mean \emph{Dice} scores and lower variance consistently across all datasets (Figure \ref{fig:boxplot3})}. Of note, \emph{SD} approach outperformed \emph{FCN}, successfully reproducing their results reported in~\cite{selected_dimensions}. 
\P{In Table~\ref{tab2} we can see a significant margin in the performance between $N$D methods like \emph{2D UNet},\emph{3D UNet},\emph{UNet++} and $3D\rightarrow 2D$ methods.} This \P{indicates} that the problem of $3D\rightarrow 2D$ segmentation cannot be solved efficiently with existing $2D$ or $3D$  methods.

\P{\textit{External test set} results are provided in Table~\ref{tab3}. The proposed method outperforms all other approaches, with significant improvements in \textit{GA2}.}  The improvement over \emph{SD}\cite{selected_dimensions} on GA 1 dataset is slightly below the significance threshold, \H{$\text{p-value}=0.063$}, but the boxplots (Figure \ref{fig:boxplot_cross}) highlight the lower variance and higher median \emph{Dice} of the proposed architecture.
\begin{table}[!htb]
    \caption{Experiments results. Left: cross-validation, right: external test set. (*: p-value$\le0.05$; **: p-value$\le10^{-5}$; ***: p-value$\le10^{-10}$).}
    \begin{minipage}{.57\linewidth}
      
\centering
\label{tab2}
\centering
\resizebox{\textwidth}{!}{\begin{tabular}{|c|l|l|l|l|l|l|}
\hline
Dataset & \multicolumn{2}{c|}{GA 1} & \multicolumn{2}{c|}{GA 2} & \multicolumn{2}{c|}{Vessel 1} \\
\hline
Model & Dice & HD95 & Dice & HD95 & Dice & HD95 \\
\hline
UNet 3D \cite{unet} & $0.71^{***}$ & $0.71^{***}$ & $0.85^{***}$ & $0.58^{***}$ & - & - \\
UNet 2D \cite{unet} & $0.73^{***}$ & $1.43^{***}$ & $0.82^{***}$ & $1.18^{***}$ & $0.51^{***}$ & $0.63^{***}$ \\
UNet++ \cite{unet} & $0.75^{***}$ & $1.22^{***}$ & $0.83^{***}$ & $1^{***}$ & $0.50^{***}$ & $0.65^{***}$ \\
\hline
FCN \cite{selected_dimensions} & $0.769^{***}$ & $0.533^{***}$ & $0.890^{***}$ & $0.114^{***}$ & $0.358^{***}$ & $0.934^{***}$ \\
IPN \cite{IPN} & $0.793^{***}$ & $0.523^{**}$ & $0.910^{***}$ & $0.116^{***}$ & $0.630^{***}$ & $0.482^{*}$\\
SD \cite{selected_dimensions}  & $0.788^{***}$ & $0.591^{***}$ & $0.919^{***}$ & $0.092^{***}$ & $0.654^{***}$ & $0.447^{**}$ \\
\hline
3D2D & $0.791^{**}$ & $0.563^{**}$ & $0.927^{***}$ & $0.070^{**}$ & $0.650^{***}$ & $0.451^{*}$ \\
\textbf{Proposed} & \textbf{0.820} & \textbf{0.336} & \textbf{0.935} & \textbf{0.052}   & \textbf{0.684} & \textbf{0.367} \\
\hline
\end{tabular}}

    \end{minipage}%
    \hfill
    \begin{minipage}{.42\linewidth}
      \centering
\label{tab3}
\resizebox{\textwidth}{!}{\begin{tabular}{|c|l|l|l|l|}
\hline
Dataset & \multicolumn{2}{c|}{GA 1} & \multicolumn{2}{c|}{GA 2}\\
\hline
Model & Dice & HD95 & Dice & HD95 \\
\hline
FCN \cite{selected_dimensions} & $0.761^{***}$ & $0.426^{***}$ & $0.878^{***}$ & $0.141^{***}$\\
IPN \cite{IPN} & $0.796^{***}$ & $0.549^{**}$ & $0.903^{***}$ & $0.094^{**}$\\
SD \cite{selected_dimensions} & $0.817$ & $0.404$ & $0.905^{***}$ & $0.113^{***}$\\
\hline
3D2D & $0.819^{*}$ & $0.382$ & $0.906^{***}$ & $0.113^{**}$\\
\textbf{Proposed} & \textbf{0.824} & \textbf{0.310} & \textbf{0.915} & \textbf{0.079} \\
\hline
\end{tabular}}

    \end{minipage} 
\end{table}


\section{Conclusion}
The problem of image segmentation in a subset of dimensions is a characteristic of a few relevant clinical applications. It can not be efficiently solved with well-studied 2D\P{-to-2D} or 3D\P{-to-3D} segmentation methods. In this paper, we first analyzed and discussed the existing approaches designed for $N\text{D}\rightarrow M\text{D}$ segmentation. For instance, existing methods share the same features, like large receptive field and encoder performing dimensionality reduction. The skip-connections, however, are implemented as subnetworks or not used at all.

Based on this analysis we proposed a novel convolutional neural network architecture for image segmentation in a subset of input dimensions. It consists of encoder that doesn't completely reduce any of the dimensions; the decoder that restores only the dimensions where the segmentation is \H{needed}; and \H{projective skip-connections} that help to link the encoder and the decoder of the network. The proposed method was tested on three medical datasets and it clearly outperformed the state of the art in two $3D\rightarrow 2D$ retinal OCT segmentation tasks.

\subsubsection*{Acknowledgements:} The financial support by the Austrian Federal Ministry for Digital and Economic Affairs, the National Foundation for Research, Technology and Development and the Christian Doppler Research Association is gratefully acknowledged.

\bibliographystyle{splncs04}
\bibliography{bibtex}
\newpage
%
\documentclass[./paper1305.tex]{subfiles}

\renewcommand{\thefootnote}{\fnsymbol{footnote}}

%

\graphicspath{ {images/} }

\newif\ifdraft
\draftfalse 

\ifdraft
\def\H#1{\textcolor{Aquamarine}{#1}}
\def\Hc#1{\textcolor{Aquamarine}{\textit{\textsf{ \small [HB: #1]}}}}  
\def\Hd#1{\textcolor{purple}{\textit{[deleted: #1]}}}  

\def\D#1{\textcolor{Orange}{#1}}
\def\Dc#1{\textcolor{Orange}{\textit{\textsf{ \small [DL: #1]}}}}  
\def\Dd#1{\textcolor{purple}{\textit{[deleted: #1]}}}  

\def\P#1{\textcolor{blue}{#1}}
\def\Pc#1{\textcolor{blue}{\textit{\textsf{ \small [PS: #1]}}}}  
\def\Pd#1{\textcolor{purple}{\textit{[deleted: #1]}}}  

\else
\def\H#1{#1}
\def\Hc#1{}  
\def\Hd#1{}  

\def\D#1{#1}
\def\Dc#1{}  
\def\Dd#1{}  

\def\P#1{#1}
\def\Pc#1{}  
\def\Pd#1{}  

\fi

\begin{document}

\begin{figure}[t]
\begin{flushleft}
\LARGE\textbf{Supplementary Materials}
\hspace{20px}

\end{flushleft}
\centering
\includegraphics[width=0.8\linewidth]{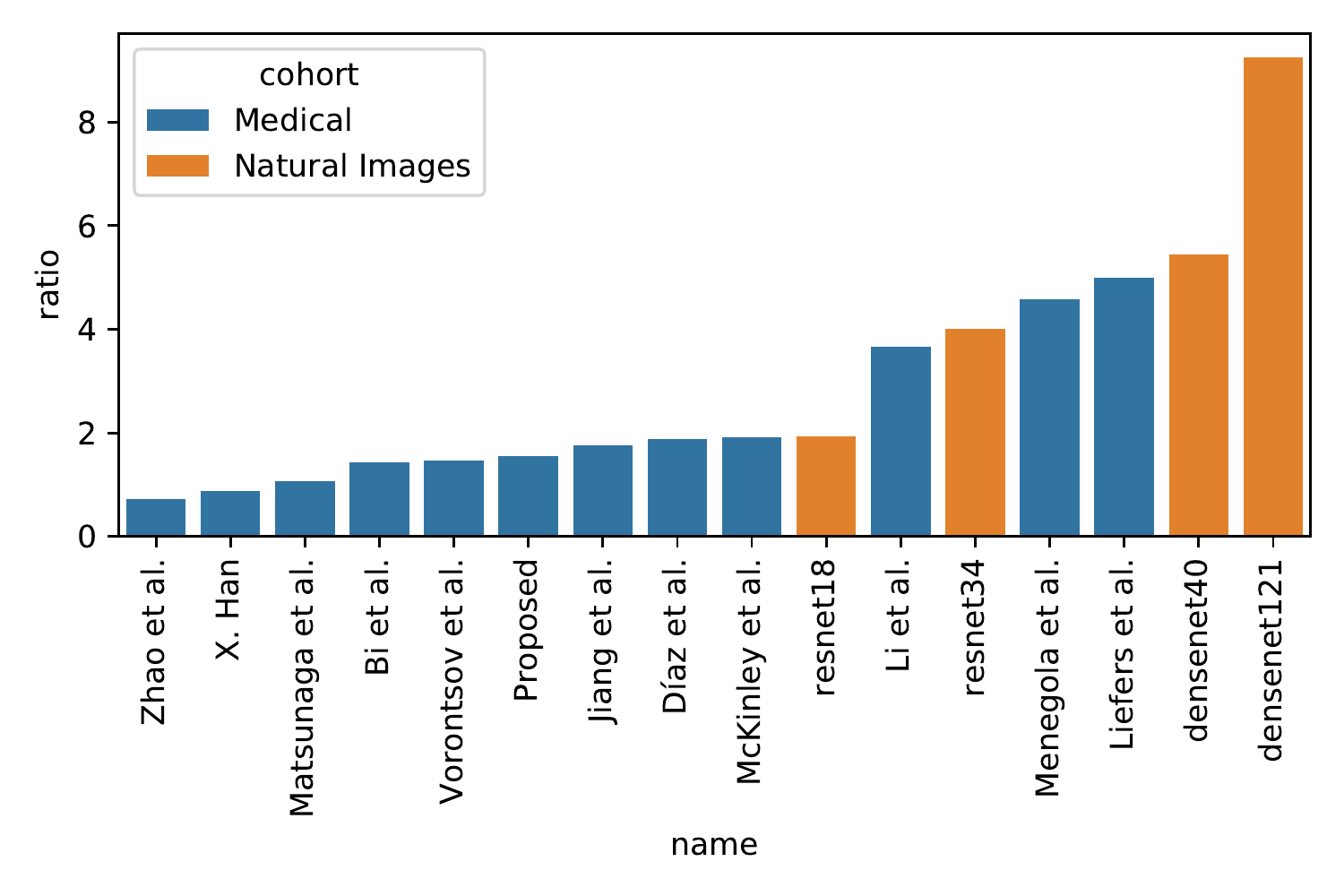}
\caption{\H{Ratio of the receptive field size and the input image size for the main CNN architectures designed for either medical or natural images.}}
\label{fig:barplot}
\end{figure}

\begin{figure}[t]
\centering
\includegraphics[width=0.8\linewidth]{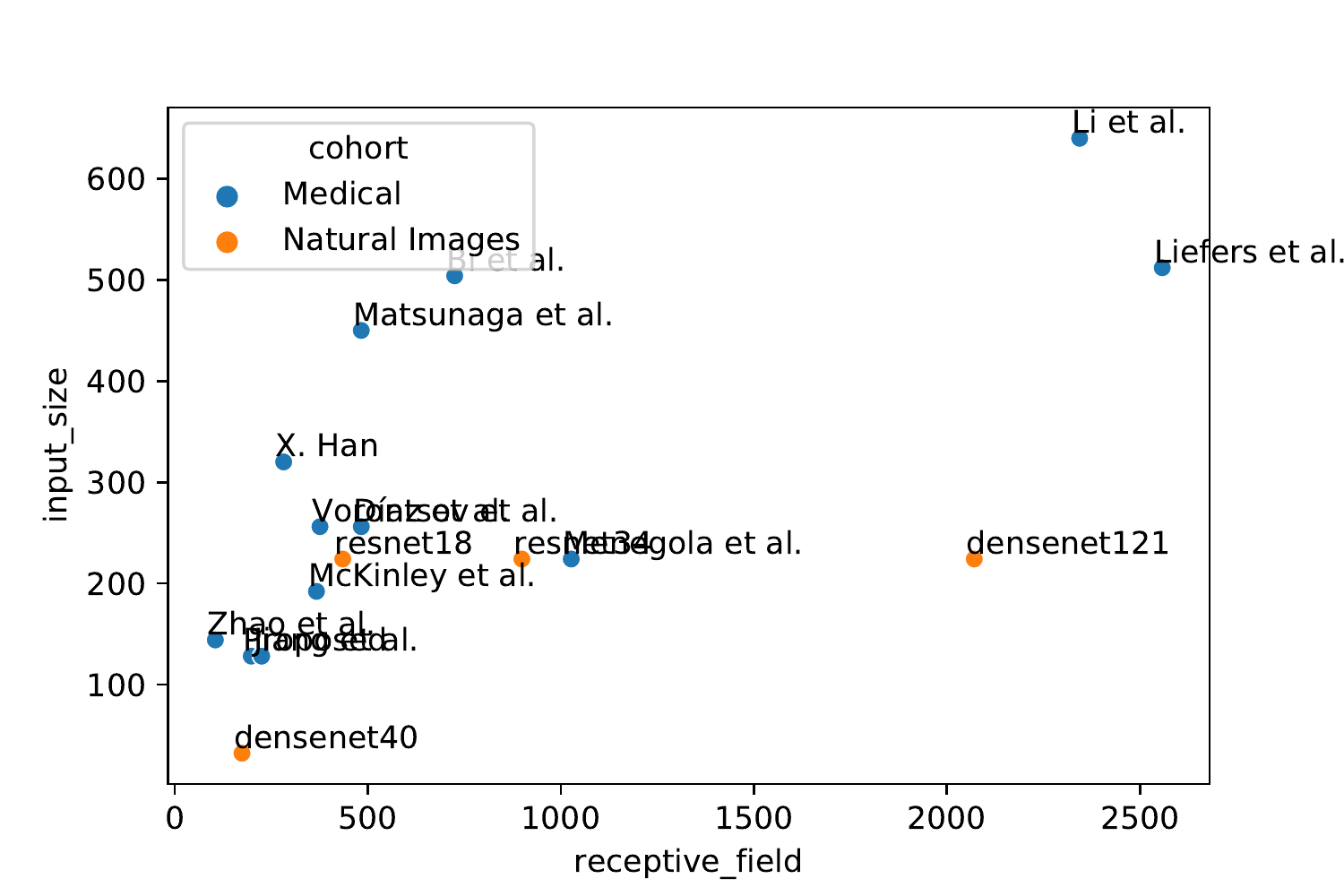}
\caption{The scatter plot of different networks where x-axis corresponds to receptive field size, and y-axis corresponds to input image size, during training.}
\label{fig:scatterplot}
\end{figure}

\begin{figure}[t]
\centering
\begin{subfigure}{.25\textwidth}
  \centering
  \includegraphics[width=\linewidth]{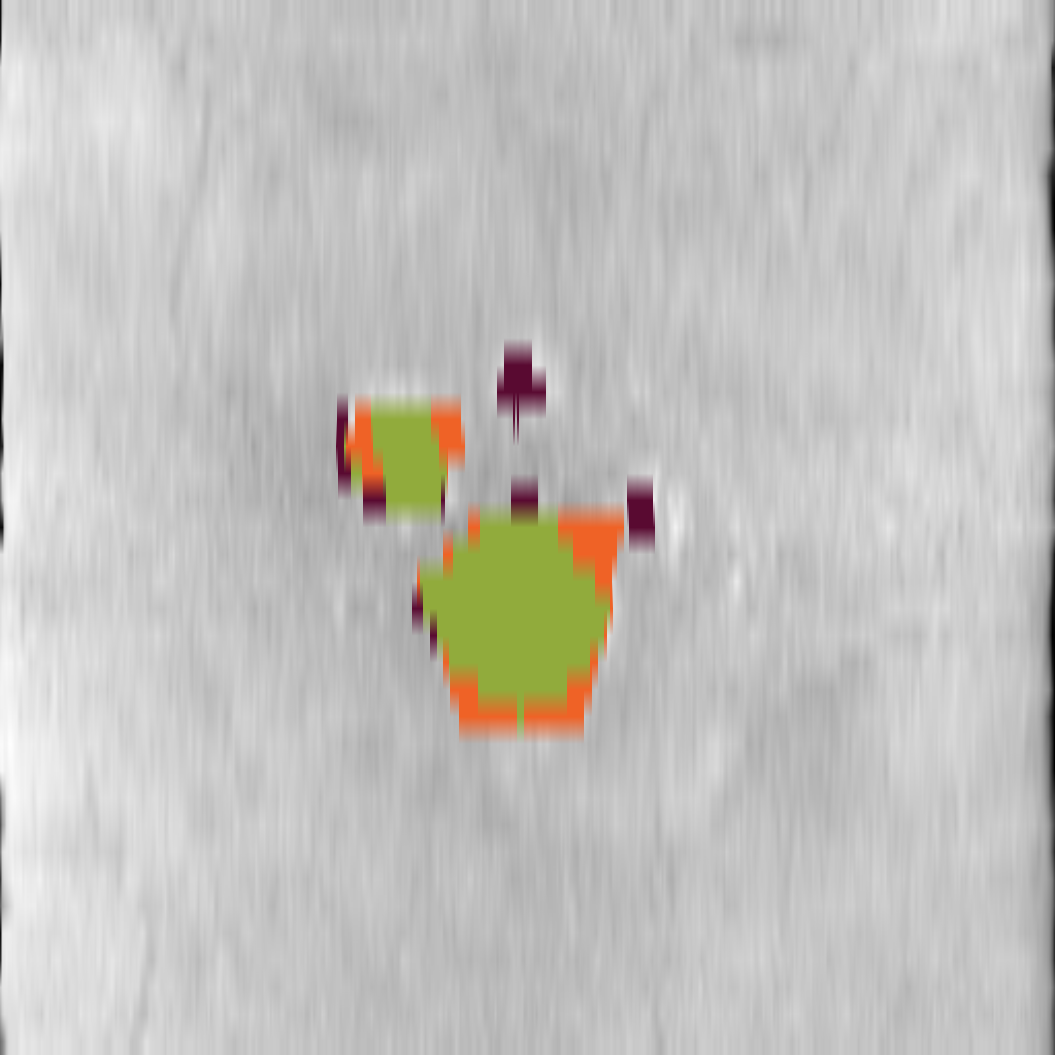}
  \caption{}
\end{subfigure}%
\begin{subfigure}{.25\textwidth}
  \centering
  \includegraphics[width=\linewidth]{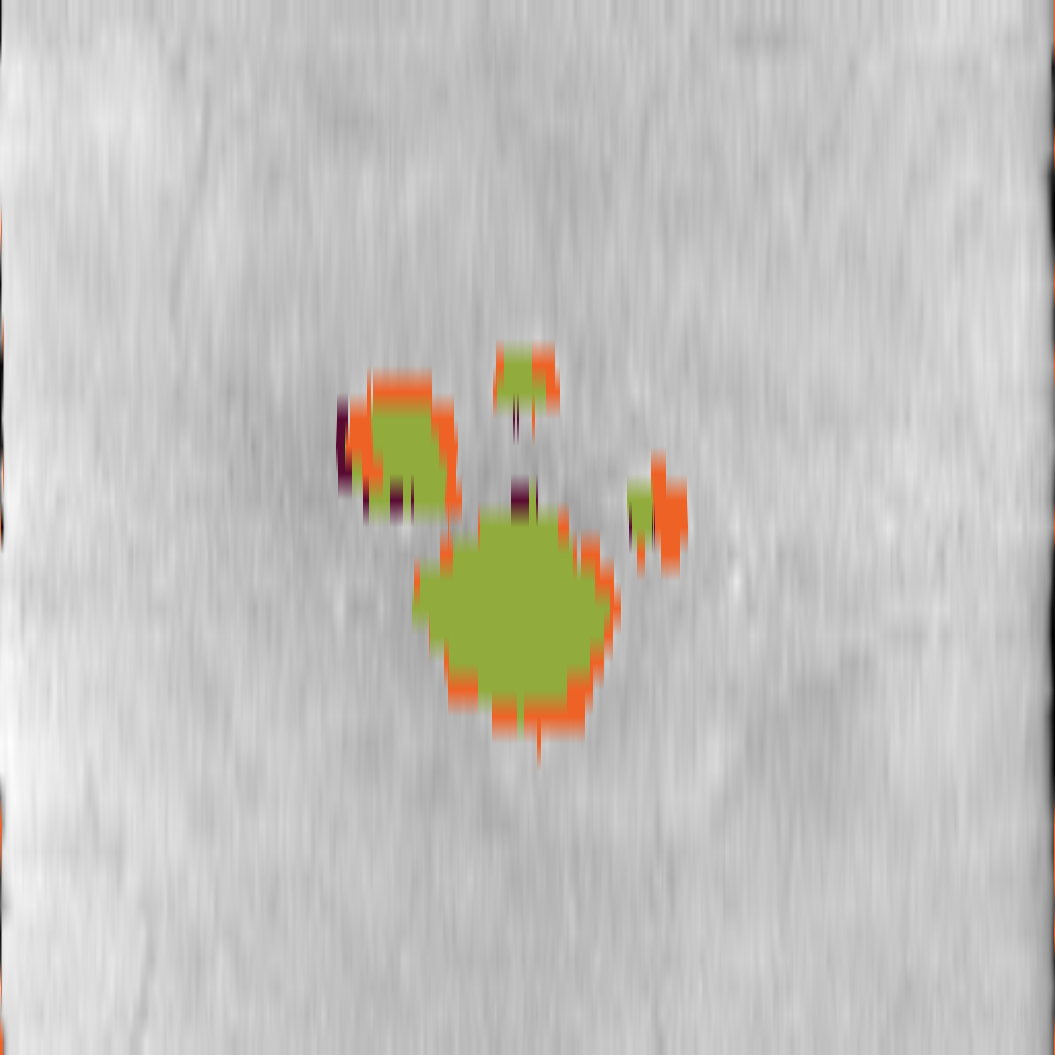}
  \caption{}
\end{subfigure}%
\begin{subfigure}{.25\textwidth}
  \centering
  \includegraphics[width=\linewidth]{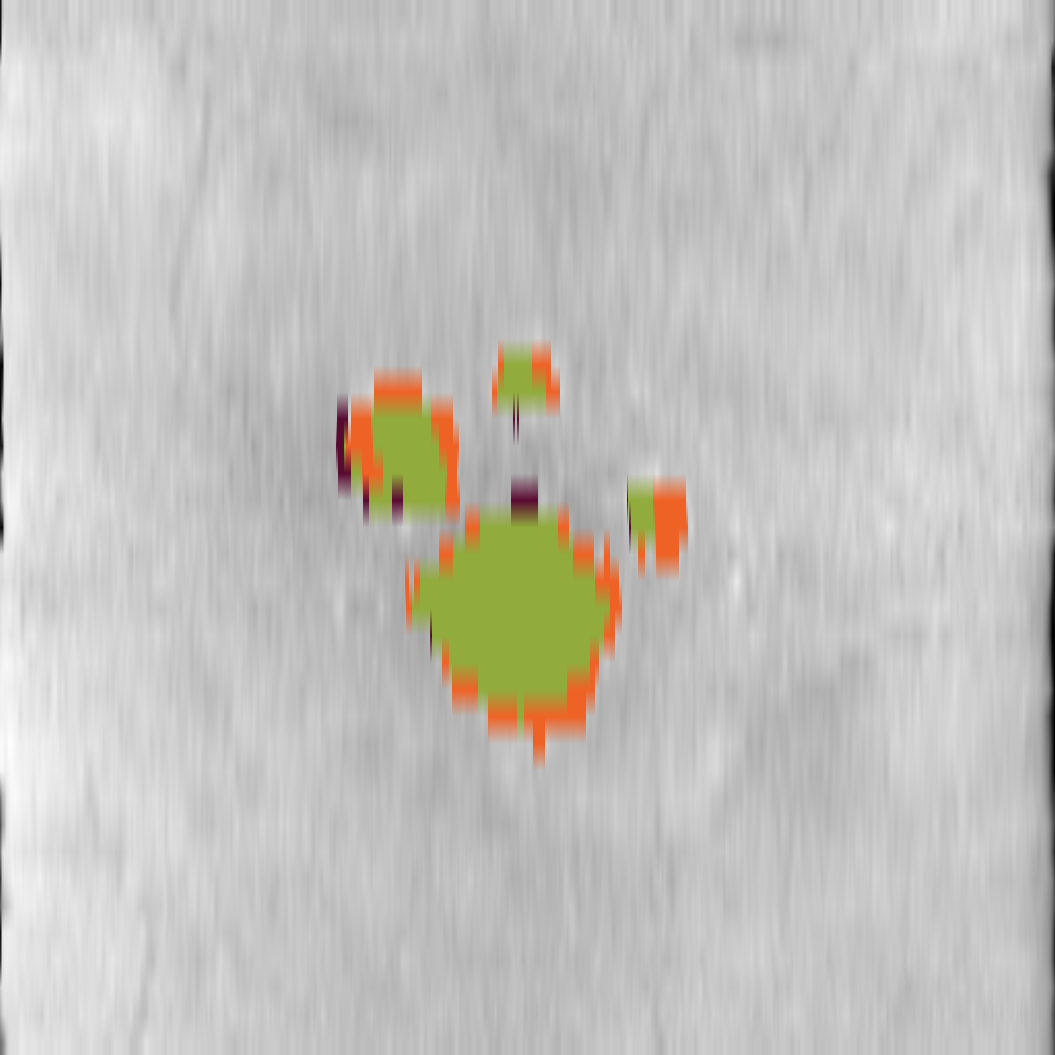}
  \caption{}
\end{subfigure}%
\begin{subfigure}{.25\textwidth}
  \centering
  \includegraphics[width=\linewidth]{gimage_proj_res_prop.png}
  \caption{}
  \label{fig:vess_example:d}
\end{subfigure}

\begin{subfigure}{.25\textwidth}
  \centering
  \includegraphics[width=\linewidth]{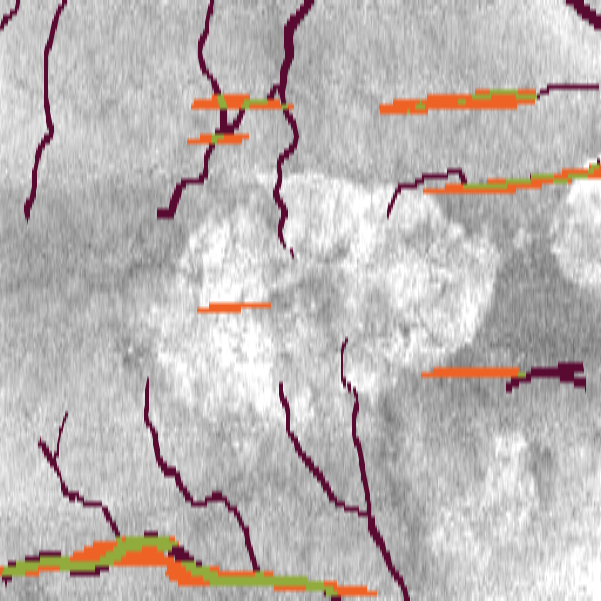}
  \caption{}
\end{subfigure}%
\begin{subfigure}{.25\textwidth}
  \centering
  \includegraphics[width=\linewidth]{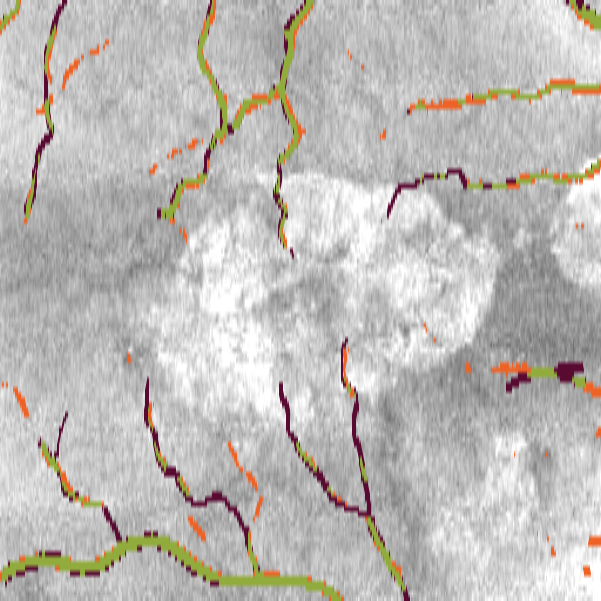}
  \caption{}
\end{subfigure}%
\begin{subfigure}{.25\textwidth}
  \centering
  \includegraphics[width=\linewidth]{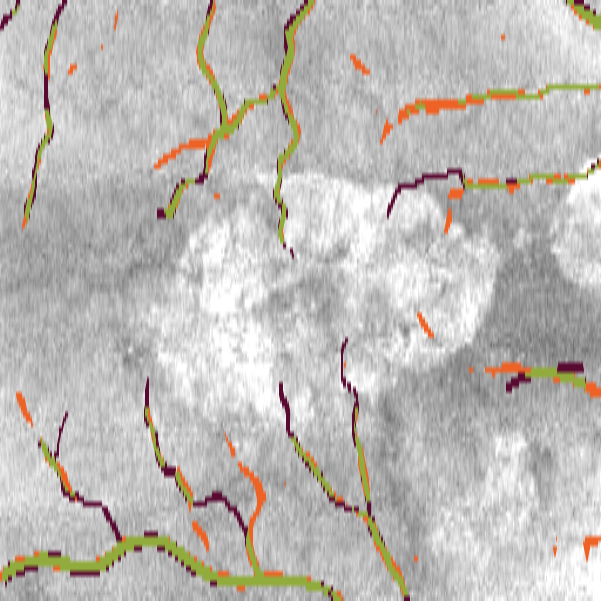}
  \caption{}
\end{subfigure}%
\begin{subfigure}{.25\textwidth}
  \centering
  \includegraphics[width=\linewidth]{vimage_proj_res_prop.png}
  \caption{}
  \label{fig:vess_example:d}
\end{subfigure}

\caption{Qualitative example of the results. Geographic Atrophy and retinal blood vessels segmentation (a),(e) FCN. (b),(f) IPN (c),(g) SD. (d),(h) Proposed method. Green region - true positives, orange - false positives, dark red - false negatives.}
\label{fig:results_example}
\end{figure}

\end{document}

\end{document}